\documentclass[twocolumn]{aastex63}
\pdfoutput=1
\usepackage{amsmath}
\usepackage{amsfonts}
\usepackage{amssymb}
\usepackage{url}

\usepackage{xspace}
\usepackage{xcolor}
\usepackage{tikz}
\usepackage{enumitem}

\usepackage[latin1]{inputenc}
\usepackage[T1]{fontenc}
\usepackage{soul}

\usepackage{bm}

\newcommand{\todo}{\ifmmode \text{\color{red}\Huge{\(\bullet\)}} \else {\color{red}{\Huge$\bullet$}}\fi}
\newcommand{\tido}{\ifmmode {{\color{red}\bullet}} \else {\color{red}$\bullet$}\fi}

\newcommand{\E        }[1]{\ifmmode 10^{#1} \else $10^{#1}$\fi}
\newcommand{\tE        }[1]{\ifmmode \times10^{#1} \else $\times10^{#1}$\fi}
\newcommand{\til}{\ifmmode \sim \else $\sim$\fi}
\renewcommand{\~} {\ifmmode \sim \else $\sim$\fi}

\newcommand{\pc}	{\ifmmode {\rm pc} \else pc\fi}
\newcommand{\kpc}	{\ifmmode {\rm kpc} \else kpc\fi}
\newcommand{\ld}	{\ifmmode {\rm l.d.} \else l.d.\fi}
\newcommand{\kms}	{\ifmmode {\rm km\,s}^{-1} \else km\,s$^{-1}$\fi}
\newcommand{\cc}	{\ifmmode {\rm cm}^{-3}    \else cm$^{-3}$\fi}
\newcommand{\cmii}	{\ifmmode {\rm cm}^{-2}    \else cm$^{-2}$\fi}
\newcommand{\ergs}	{\ifmmode {\rm erg\,s}^{-1} \else erg s$^{-1}$\fi}
\newcommand{\ergcms}	{\ifmmode {\rm erg\,cm}^{-2}\,{\rm s}^{-1} \else erg\,cm$^{-2}$\,s$^{-1}$\fi}
\newcommand{\ergcmsA}	{\ifmmode {\rm erg\,cm}^{-2}\,{\rm s}^{-1}\,{\rm\AA}^{-1}
\else erg\,cm$^{-2}$\,s$^{-1}$\,\AA$^{-1}$\fi}
\newcommand{  \ergcmsHz  }{\ifmmode{\rm erg\,cm}^{-2}\,{\rm s}^{-1}\,{\rm Hz}^{-1}
                       \else ergs\,cm$^{-2}$\,s$^{-1}$\,Hz$^{-1}$\fi}
\newcommand{\kev}	{\ifmmode {\rm keV} \else keV\fi}

\newcommand{\mic}	{\ifmmode {\rm \mu m} \else $\mu$m\fi}
\newcommand{\vFWHM}	{\ifmmode v_{\mbox{\tiny FWHM}} \else $v_{\mbox{\tiny FWHM}}$\fi}
\newcommand{\vBLR}	{\ifmmode v_{\mbox{\tiny BLR}} \else $v_{\mbox{\tiny BLR}}$\fi}
\newcommand{\sigBLR}	{\ifmmode \sigma_{\mbox{\tiny BLR}} \else $\sigma_{\mbox{\tiny BLR}}$\fi}
\newcommand{\vNLR}	{\ifmmode v_{\mbox{\tiny NLR}} \else $v_{\mbox{\tiny NLR}}$\fi}
\newcommand{\tauBLR}	{\ifmmode \tau_{\mbox{\tiny BLR}} \else $\tau_{\mbox{\tiny BLR}}$\fi}

\newcommand{\Hubble}	{\ifmmode {\rm km\,s}^{-1}\,{\rm Mpc}^{-1} \else km\,s$^{-1}$\,Mpc$^{-1}$\fi}
\newcommand{\NDunit}	{\ifmmode {\rm Mpc}^{-3} \else Mpc$^{-3}$\fi}
\newcommand{\LFunit}	{\ifmmode {\rm Mpc}^{-3}\,{\rm mag}^{-1} \else Mpc$^{-3}$\,mag$^{-1}$\fi}
\newcommand{\MFunit}	{\ifmmode {\rm Mpc}^{-3}\,{\rm dex}^{-1} \else Mpc$^{-3}$\,dex$^{-1}$\fi}

\newcommand{\Msun}{\ifmmode M_{\odot} \else $M_{\odot}$\fi}
\newcommand{\Lsun}{\ifmmode L_{\odot} \else $L_{\odot}$\fi}
\newcommand{\Zsun}{\ifmmode Z_{\odot} \else $Z_{\odot}$\fi}
\newcommand{\mpyr}{\ifmmode \Msun\,{\rm yr}^{-1} \else $\Msun\,{\rm yr}^{-1}$\fi}

\newcommand{\Msol}{\Msun}

\newcommand{\qnote}{\ifmmode q_{0} \else $q_{0}$\fi}
\newcommand{\Hnote}{\ifmmode H_{0} \else $H_{0}$\fi}
\newcommand{\hnote}{\ifmmode h_{0} \else $h_{0}$\fi}
\newcommand{\anote}{\ifmmode a_{0} \else $a_{0}$\fi}
\newcommand{\tnote}{\ifmmode t_{0} \else $t_{0}$\fi}

\def\gsim{\;\rlap{\lower 2.5pt \hbox{$\sim$}}\raise 1.5pt\hbox{$>$}\;}
\def\lsim{\;\rlap{\lower 2.5pt \hbox{$\sim$}}\raise 1.5pt\hbox{$<$}\;}

\newcommand{  \Halpha   }{\ifmmode {\rm H}\alpha \else H$\alpha$\fi}

\newcommand{  \ha       }{\Halpha}
\newcommand{  \Hbeta    }{\ifmmode {\rm H}\beta \else H$\beta$\fi}
\newcommand{  \hbeta    }{\Hbeta}
\newcommand{  \hb       }{\Hbeta}
\newcommand{  \Hgamma   }{\ifmmode {\rm H}\gamma \else H$\gamma$\fi}
\newcommand{  \Hdelta   }{\ifmmode {\rm H}\delta \else H$\delta$\fi}
\newcommand{  \Lya      }{\ifmmode {\rm Ly}\alpha \else Ly$\alpha$\fi}
\newcommand{  \Lyb      }{\ifmmode {\rm Ly}\beta \else Ly$\beta$\fi}
\newcommand{  \Pa       }{\ifmmode {\rm P}\alpha \else P$\alpha$\fi}
\newcommand{  \Pb       }{\ifmmode {\rm P}\beta \else P$\beta$\fi}
\newcommand{  \Bra      }{\ifmmode {\rm Br}\alpha \else Br$\alpha$\fi}
\newcommand{  \Brg      }{\ifmmode {\rm Br}\gamma \else Br$\gamma$\fi}
\newcommand{  \hii      }{\ifmmode {\rm H}\,\textsc{ii} \else H\,\textsc{ii}\fi}
\newcommand{  \hei      }{\ifmmode {\rm He}\,\textsc{i} \else He\,\textsc{i}\fi}
\newcommand{  \heii     }{\ifmmode {\rm He}\,\textsc{ii} \else He\,\textsc{ii}\fi}
\newcommand{  \HeIIuv   }{\ifmmode {\rm He}\,\textsc{ii}\,\lambda1640 \else He\,\textsc{ii}\,$\lambda1640$\fi}
\newcommand{  \HeIIop   }{\ifmmode {\rm He}\,\textsc{ii}\,\lambda4686 \else He\,\textsc{ii}\,$\lambda4686$\fi}
\newcommand{  \CII	}{\ifmmode \left[{\rm C}\,\textsc{ii}\right]\,\lambda157.74\,\mu{\rm m} \else [C\,{\sc ii}]\ $\lambda157.74\,\mu{\rm m}$\fi}
\newcommand{  \cii	}{\ifmmode \left[{\rm C}\,\textsc{ii}\right] \else [C\,{\sc ii}]\fi}

\newcommand{  \ciii     }{\ifmmode {\rm C}\,\textsc{iii}\right] \else C\,\textsc{iii}]\fi}
\newcommand{  \CIII     }{\ifmmode {\rm C}\,\textsc{iii}\right]\,\lambda1909 \else C\,\textsc{iii}]\,$\lambda1909$\fi}
\newcommand{  \civ      }{\ifmmode {\rm C}\,\textsc{iv}  \else C\,\textsc{iv}\fi}
\newcommand{  \CIV      }{\ifmmode {\rm C}\,\textsc{iv}\,\lambda1549 \else C\,\textsc{iv}\,$\lambda1549$\fi}
\newcommand{  \NIIopt   }{\ifmmode \left[{\rm N}\,\textsc{ii}\right]\,\lambda6584 \else [N\,\textsc{ii}]\,$\lambda6584$\fi}
\newcommand{  \nii      }{\ifmmode \left[{\rm N}\,\textsc{ii}\right]  \else [N\,\textsc{ii}]\fi}
\newcommand{  \niii     }{\ifmmode {\rm N}\,\textsc{iii} \else N\,\textsc{iii}\fi}
\newcommand{  \NIII     }{\ifmmode {\rm N}\,\textsc{iii}\,\lambda4640 \else N\,\textsc{iii}\,$\lambda4640$\fi}
\newcommand{  \niv      }{\ifmmode {\rm N}\,\textsc{iv}  \else N\,\textsc{iv}\fi}
\newcommand{  \NIVuv    }{\ifmmode {\rm N}\,\textsc{iv}\,\lambda1486 \else N\,\textsc{iv}\,$\lambda1486$\fi}
\newcommand{  \nv       }{\ifmmode {\rm N}\,\textsc{v}   \else N\,\textsc{v}\fi}
\newcommand{\oi}{\ifmmode \left[{\rm O}\,\textsc{i}\right] \else [O\,{\sc i}]\fi}
\newcommand{\OI}{\ifmmode \left[{\rm O}\,\textsc{i}\right]\,\lambda6300 \else [O\,{\sc i}]$\,\lambda6300$\fi}
\newcommand{\oii}{\ifmmode \left[{\rm O}\,\textsc{ii}\right] \else [O\,{\sc ii}]\fi}
\newcommand{\OII}{\ifmmode \left[{\rm O}\,\textsc{ii}\right]\,\lambda3727 \else [O\,{\sc ii}]\,$\lambda3727$\fi}
\newcommand{\oiii}{\ifmmode \left[{\rm O}\,\textsc{iii}\right] \else [O\,{\sc iii}]\fi}
\newcommand{\OIII}{\ifmmode \left[{\rm O}\,\textsc{iii}\right]\,\lambda5007 \else [O\,{\sc iii}]\,$\lambda5007$\fi}
\newcommand{  \OIIIbf   }{\ifmmode {\rm O}\,\textsc{iii}\,\lambda3133 \else O\,\textsc{iii}\,$\lambda3133$\fi}
\newcommand{  \OIIIuv   }{\ifmmode {\rm O}\,\textsc{iii}\,\lambda1663 \else O\,\textsc{iii}\,$\lambda1663$\fi}
\newcommand{  \oiv      }{\ifmmode {\rm O}\,\textsc{iv}  \else O\,\textsc{iv}\fi}
\newcommand{  \OIVuv    }{\ifmmode {\rm O}\,\textsc{iv}\,\lambda1402  \else O\,\textsc{iv}\,$\lambda1402$\fi}
\newcommand{  \OIVIR    }{\ifmmode {\rm O}\,\textsc{iv}\,25.9\,\mu {\rm m} \else O\,\textsc{iv}\,$25.9\,\mu$m\fi}
\newcommand{  \ovi      }{\ifmmode {\rm O}\,\textsc{vi}   \else O\,\textsc{vi}\fi}
\newcommand{  \Ovi      }{\ifmmode {\rm O}\,\textsc{vi}\,\lambda1035 \else O\,\textsc{vi}\,$\lambda1035$\fi}
\newcommand{  \nei      }{\ifmmode {\rm Ne}\,\textsc{i}   \else Ne\,\textsc{i}\fi}
\newcommand{  \neii     }{\ifmmode {\rm Ne}\,\textsc{ii}  \else Ne\,\textsc{ii}\fi}
\newcommand{  \NeiiIR   }{\ifmmode {\rm Ne}\,\textsc{ii}\,12.8\,\mu {\rm m} \else Ne\,\textsc{ii}\,$12.8\,\mu$m\fi}
\newcommand{  \neiii    }{\ifmmode {\rm Ne}\,\textsc{iii} \else Ne\,\textsc{iii}\fi}
\newcommand{  \neiv     }{\ifmmode {\rm Ne}\,\textsc{iv}  \else Ne\,\textsc{iv}\fi}
\newcommand{  \nev      }{\ifmmode \left[{\rm Ne}\,\textsc{v}\right]   \else [Ne\,\textsc{v}]\fi}
\newcommand{  \NevIR    }{\ifmmode \left[{\rm Ne}\,\textsc{v}\right]\,\lambda24.3\,\mu {\rm m} \else Ne\,\textsc{v}\,$\lambda24.3\,\mu$m\fi}
\newcommand{  \nevi     }{\ifmmode {\rm Ne}\,\textsc{vi}  \else Ne\,\textsc{vi}\fi}
\newcommand{  \mgi      }{\ifmmode {\rm Mg}\,\textsc{i} \else Mg\,\textsc{i}\fi}
\newcommand{  \mgii     }{\ifmmode {\rm Mg}\,\textsc{ii} \else Mg\,\textsc{ii}\fi}
\newcommand{  \MgII     }{\ifmmode {\rm Mg}\,\textsc{ii}\,\lambda2798 \else Mg\,\textsc{ii}\,$\lambda2798$\fi}
\newcommand{  \sii      }{\ifmmode {\rm S}\,\textsc{ii} \else S\,\textsc{ii}\fi}
\newcommand{  \siii     }{\ifmmode {\rm S}\,\textsc{iii} \else S\,\textsc{iii}\fi}
\newcommand{  \siv      }{\ifmmode {\rm S}\,\textsc{iv} \else S\,\textsc{iv}\fi}
\newcommand{  \sili     }{\ifmmode {\rm Si}\,\textsc{i}   \else Si\,\textsc{i}\fi}
\newcommand{  \silii    }{\ifmmode {\rm Si}\,\textsc{ii}  \else Si\,\textsc{ii}\fi}
\newcommand{  \Siliv    }{\ifmmode {\rm Si}\,\textsc{iv}  \else Si\,\textsc{iv}\fi}
\newcommand{  \SilIVuv  }{\ifmmode {\rm Si}\,\textsc{iv}\,\lambda1400  \else Si\,\textsc{iv}\,$\lambda1400$\fi}
\newcommand{  \AlIII   }{\ifmmode {\rm Al}\,\textsc{iii}\,\lambda1857 \else Al\,\textsc{iii}\,$\lambda1857$\fi}
\newcommand{  \Aliii   }{\ifmmode {\rm Al}\,\textsc{iii} \else Al\,\textsc{iii}\fi}
\newcommand{  \caii     }{\ifmmode {\rm Ca}\,\textsc{ii} \else Ca\,\textsc{ii}\fi}
\newcommand{  \feii     }{\ifmmode {\rm Fe}\,\textsc{ii} \else Fe\,\textsc{ii}\fi}
\newcommand{  \feiii    }{\ifmmode {\rm Fe}\,\textsc{iii} \else Fe\,\textsc{iii}\fi}
\newcommand{  \Kalpha   }{\ifmmode {\rm K}\alpha \else K$\alpha$\fi}

\newcommand{ \Lhb   }{\ifmmode L_{\hb} \else $L_{\hb}$\fi}
\newcommand{ \Lha   }{\ifmmode L_{\ha} \else $L_{\ha}$\fi}
\newcommand{ \fwhb  }{\ifmmode {\rm FWHM}\left(\hb\right) \else FWHM(\hb)\fi}
\newcommand{\sighb  }{\ifmmode \sigma\left(\hb\right) \else $\sigma\left(\hb\right)$\fi}
\newcommand{ \ewhb  }{\ifmmode {\rm EW}\left(\hb\right) \else EW(\hb)\fi}
\newcommand{ \fwha  }{\ifmmode {\rm FWHM}\left(\ha\right) \else FWHM(\ha)\fi}
\newcommand{ \ewha  }{\ifmmode {\rm EW}\left(\ha\right) \else EW(\ha)\fi}
\newcommand{ \Lmg   }{\ifmmode L\left(\mgii\right) \else $L\left(\mgii\right)$\fi}
\newcommand{ \fwmg  }{\ifmmode {\rm FWHM}\left(\mgii\right) \else FWHM(\mgii)\fi}
\newcommand{ \Lciv  }{\ifmmode L\left(\civ\right) \else $L\left(\civ\right)$\fi}
\newcommand{ \fwciv }{\ifmmode {\rm FWHM}\left(\civ\right) \else FWHM(\civ)\fi}
\newcommand{ \fwhm  }{\ifmmode {\rm FWHM} \else FWHM\fi} 
\newcommand{ \voff  }{\ifmmode v_{\rm off} \else $v_{\rm off}$\fi} 
\newcommand{ \vmax  }{\ifmmode v_{\rm max} \else $v_{\rm max}$\fi} 

\newcommand{ \mumg  }{\ifmmode \mu\left(\mgii\right) \else $\mu\left(\mgii\right)$\fi}
\newcommand{ \fmg   }{\ifmmode f\left(\mgii\right) \else $f\left(\mgii\right)$\fi}
\newcommand{ \muciv }{\ifmmode \mu\left(\civ\right) \else $\mu\left(\civ\right)$\fi}
\newcommand{ \fciv  }{\ifmmode f\left(\civ\right) \else $f\left(\civ\right)$\fi}

\newcommand{  \auvo     }{\ifmmode \alpha_{\nu,{\rm UVO}} \else $\alpha_{\nu,{\rm UVO}}$\fi}
\newcommand{  \Ledd     }{\ifmmode L_{\rm Edd} \else $L_{\rm Edd}$\fi}
\newcommand{  \lamLlam  }{\ifmmode \lambda L_{\lambda} \else $\lambda L_{\lambda}$\fi}
\newcommand{  \lLl      }{\ifmmode \lambda L_{\lambda} \else $\lambda L_{\lambda}$\fi}
\newcommand{  \nuLnu    }{\ifmmode \nu L_{\nu} \else $\nu L_{\nu}$\fi}
\newcommand{  \nLn      }{\ifmmode \nu L_{\nu} \else $\nu L_{\nu}$\fi}
\newcommand{  \Luv      }{\ifmmode L_{1450} \else $L_{1450}$\fi}
\newcommand{  \Lop      }{\ifmmode L_{5100} \else $L_{5100}$\fi}
\newcommand{  \lLop     }{\ifmmode \log\left(\Lop/\ergs\right) \else $\log\left(\Lop/\ergs\right)$\fi}
\newcommand{  \Lthree   }{\ifmmode L_{3000} \else $L_{3000}$\fi}
\newcommand{  \lLthree  }{\ifmmode \log\left(\Lthree/\ergs\right) \else $\log\left(\Lthree/\ergs\right)$\fi}
\newcommand{  \Lsix      }{\ifmmode L_{6200} \else $L_{6200}$\fi}
\newcommand{  \lLisx     }{\ifmmode \log\left(\Lop/\ergs\right) \else $\log\left(\Lop/\ergs\right)$\fi}
\newcommand{  \Lhard    }{\ifmmode L_{\rm 2-10} \else $L_{\rm 2-10}$\fi}
\newcommand{  \Fhard    }{\ifmmode F_{\rm 2-10} \else $F_{\rm 2-10}$\fi}
\newcommand{  \Lsoft    }{\ifmmode L_{\rm 0.5-2} \else $L_{\rm 0.5-2}$\fi}
\newcommand{  \Luhard    }{\ifmmode L_{\rm 14-195} \else $L_{\rm 14-195}$\fi}
\newcommand{  \Fuhard    }{\ifmmode F_{\rm 14-195} \else $F_{\rm 14-195}$\fi}
\newcommand{  \Lxray    }{\ifmmode L_{\rm X} \else $L_{\rm X}$\fi}
\newcommand{  \Fxray    }{\ifmmode F_{\rm X} \else $F_{\rm X}$\fi}

\newcommand{\Fthree}{\ifmmode F_{3000} \else $F_{3000}$\fi}
\newcommand{\fuv}{\ifmmode f_{\lambda}\left(1450{\rm \AA}\right) \else $f_{\lambda}\left(1450 {\rm \AA}\right)$\fi}
\newcommand{\fthree}{\ifmmode f_{\lambda}\left(3000{\rm \AA}\right) \else $f_{\lambda}\left(3000{\rm \AA}\right)$\fi}
\newcommand{\fH}{\ifmmode f_{\lambda}\left(1.65\micron\right) \else
$f_{\lambda}\left(1.65\micron\right)$\fi}

\newcommand{\fbol}{\ifmmode f_{\rm bol} \else $f_{\rm bol}$\fi}
\newcommand{\fbolwv}{\ifmmode f_{\rm bol}\left(\lambda\right) \else $f_{\rm bol}\left(\lambda\right)$\fi}
\newcommand{\fbolopt}{\ifmmode f_{\rm bol}\left(5100{\rm \AA}\right) \else $f_{\rm bol}\left(5100{\rm \AA}\right)$\fi}
\newcommand{\fbolthree}{\ifmmode f_{\rm bol}\left(3000{\rm \AA}\right) \else $f_{\rm bol}\left(3000{\rm \AA}\right)$\fi}
\newcommand{\fboluv}{\ifmmode f_{\rm bol}\left(1450{\rm \AA}\right) \else $f_{\rm bol}\left(1450{\rm \AA}\right)$\fi}

\newcommand{\fbolbat}{\ifmmode f_{\rm bol}\left(14-150\,\kev\right) \else $f_{\rm bol}\left(14-150\,\kev\right)$\fi}
\newcommand{\fbolhard}{\ifmmode f_{\rm bol}\left(2-10\,\kev\right) \else $f_{\rm bol}\left(2-10\,\kev\right)$\fi}

\newcommand{\fobs}{\ifmmode f_{\rm obs} \else $f_{\rm obs}$\fi}

\newcommand{  \mbh      }{\ifmmode M_{\rm BH} \else $M_{\rm BH}$\fi}
\newcommand{  \lmbh     }{\ifmmode \log\left(\mbh/\Msun\right) \else $\log\left(\mbh/\Msun\right)$\fi} 
\newcommand{  \lledd    }{\ifmmode L/L_{\rm Edd} \else $L/L_{\rm Edd}$\fi}
\newcommand{  \mmedd    }{\ifmmode \dot{m}/\dot{m}_{\rm \,Edd} \else $\dot{m}/\dot{m}_{\rm \,Edd}$\fi}
\newcommand{  \Lbol     }{\ifmmode L_{\rm bol} \else $L_{\rm bol}$\fi}
\newcommand{  \lbol     }{\ifmmode L_{\rm bol} \else $L_{\rm bol}$\fi}
\newcommand{  \lLbol    }{\ifmmode \log\left(\Lbol/\ergs\right) \else $\log\left(\Lbol/\ergs\right)$\fi} 
\newcommand{  \Lagn     }{\ifmmode L_{\rm AGN} \else $L_{\rm AGN}$\fi}
\newcommand{  \lagn     }{\ifmmode L_{\rm AGN} \else $L_{\rm AGN}$\fi}

\newcommand{  \tgrow     }{\ifmmode t_{\rm growth} \else $t_{\rm growth}$\fi}
\newcommand{  \tAD     }{\ifmmode t_{\rm acc} \else $t_{\rm acc}$\fi}
\newcommand{  \tacc    }{\ifmmode t_{\rm acc} \else $t_{\rm acc}$\fi}
\newcommand{  \tUni      }{\ifmmode t_{\rm Universe} \else $t_{\rm Universe}$\fi}

\newcommand{  \Mdotin	}{\ifmmode \dot{M}_{\rm infall} \else $\dot{M}_{\rm infall}$\fi}
\newcommand{  \Mdotbh	}{\ifmmode \dot{M}_{\rm BH} \else $\dot{M}_{\rm BH}$\fi}
\newcommand{  \Mdotad	}{\ifmmode \dot{M}_{\rm AD} \else $\dot{M}_{\rm AD}$\fi}
\newcommand{  \Mdotacc	}{\ifmmode \dot{M}_{\rm acc} \else $\dot{M}_{\rm acc}$\fi}
\newcommand{  \Mdotthin	}{\ifmmode \dot{M}_{\rm thin} \else $\dot{M}_{\rm thin}$\fi}
\newcommand{  \Mdotdisk	}{\ifmmode \dot{M}_{\rm disk} \else $\dot{M}_{\rm disk}$\fi}

\newcommand{  \Mindot	}{\ifmmode \dot{M}_{\rm infall} \else $\dot{M}_{\rm infall}$\fi}
\newcommand{  \Mbhdot	}{\ifmmode \dot{M}_{\rm BH} \else $\dot{M}_{\rm BH}$\fi}
\newcommand{  \Maddot	}{\ifmmode \dot{M}_{\rm AD} \else $\dot{M}_{\rm AD}$\fi}
\newcommand{  \Maccdot	}{\ifmmode \dot{M}_{\rm acc} \else $\dot{M}_{\rm acc}$\fi}
\newcommand{  \Mthdot	}{\ifmmode \dot{M}_{\rm thin} \else $\dot{M}_{\rm thin}$\fi}
\newcommand{  \Mdsdot	}{\ifmmode \dot{M}_{\rm disk} \else $\dot{M}_{\rm disk}$\fi}

\newcommand{  \as	}{\ifmmode a_{\rm *} \else $a_{\rm *}$\fi}
\newcommand{  \avec	}{\ifmmode \vec{a}_{\rm *} \else $\vec{a}_{\rm *}$\fi}
\newcommand{  \re	}{\ifmmode \eta      	 \else $\eta$\fi}
\newcommand{  \RISCO	}{\ifmmode R_{\rm ISCO}  \else $R_{\rm ISCO}$\fi}

\newcommand{  \mseed    }{\ifmmode M_{\rm seed} \else $M_{\rm seed}$\fi}
\newcommand{  \mbul     }{\ifmmode M_{\rm bulge} \else $M_{\rm bulge}$\fi} 
\newcommand{  \mstar    }{\ifmmode M_{*} \else $M_{*}$\fi} 
\newcommand{  \mgal     }{\ifmmode M_{*} \else $M_{*}$\fi} 
\newcommand{  \mhost    }{\ifmmode M_{\rm host} \else $M_{\rm host}$\fi}
\newcommand{  \mmsmall  }{\ifmmode M_{\rm BH}/M_{*} \else $M_{\rm BH}/M_{*}$\fi}
\newcommand{  \mmlarge  }{\ifmmode M_{*}/M_{\rm BH} \else $M_{*}/M_{\rm BH}$\fi}
\newcommand{  \mm       }{\mmsmall}
\newcommand{  \mmdotlarge}{\ifmmode \dot{M}_*/\Mbhdot \else $\dot{M}_*/\Mbhdot$\fi}
\newcommand{  \mmdotsmall}{\ifmmode \Mbhdot/\dot{M}_* \else $\Mbhdot/\dot{M}_*$\fi}

\newcommand{  \mmwp     }{\ifmmode \left(M_{*}/M_{\rm BH}\right) \else $\left(M_{*}/M_{\rm BH}\right)$\fi}
\newcommand{  \ml       }{\ifmmode M_{*}/L_{*} \else $M_{*}/L_{*}$\fi}
\newcommand{  \mlwp     }{\ifmmode \left(M_{*}/L\right) \else $\left(M_{*}/L\right)$\fi}
\newcommand{  \mlk      }{\ifmmode \left(M_{*}/L_{K}\right) \else $\left(M_{*}/L_{K}\right)$\fi}
\newcommand{  \sigs     }{\ifmmode \sigma_{*} \else $\sigma_{*}$\fi}
\newcommand{  \Reff     }{\ifmmode R_{\rm e} \else $R_{\rm e}$\fi}
\newcommand{  \Rvir     }{\ifmmode R_{\rm vir} \else $R_{\rm vir}$\fi}
\newcommand{  \Rtwo     }{\ifmmode R_{200} \else $R_{200}$\fi}
\newcommand{  \Rfive    }{\ifmmode R_{500} \else $R_{500}$\fi}
\newcommand{  \Rgrp     }{\ifmmode R_{\rm grp} \else $R_{\rm grp}$\fi}
\newcommand{  \nser     }{\ifmmode n_{\rm s} \else $n_{\rm s}$\fi}
\newcommand{  \LSF      }{\ifmmode L_{\rm SF}  \else $L_{\rm SF}$\fi}
\newcommand{  \LFIR     }{\ifmmode L_{\rm FIR} \else $L_{\rm FIR}$\fi}
\newcommand{  \Lfir     }{\ifmmode L_{\rm FIR} \else $L_{\rm FIR}$\fi}
\newcommand{  \LTIR     }{\ifmmode L_{\rm TIR} \else $L_{\rm TIR}$\fi}
\newcommand{  \Ltir     }{\ifmmode L_{\rm TIR} \else $L_{\rm TIR}$\fi}

\newcommand{  \mdyn     }{\ifmmode M_{\rm dyn} \else $M_{\rm dyn}$\fi} 
\newcommand{  \mgas     }{\ifmmode M_{\rm gas} \else $M_{\rm gas}$\fi} 
\newcommand{  \mh       }{\ifmmode M_{\rm h} \else $M_{\rm h}$\fi}
\newcommand{  \mhalo    }{\ifmmode M_{\rm halo} \else $M_{\rm halo}$\fi}
\newcommand{  \sfr      }{\ifmmode {\rm SFR} \else SFR\fi}

\newcommand{ \Lcii     }{\ifmmode L_{\cii} \else $L_{\cii}$\fi}
\newcommand{ \fwcii  }{\ifmmode {\rm FWHM}\cii \else FWHM\cii\fi}

\newcommand{\bj}{\ifmmode b_{\rm J} \else $b_{\rm J}$\fi}

\newcommand{\iab}{\ifmmode i_{\rm AB} \else $i_{\rm AB}$\fi}

\newcommand{\jab}{\ifmmode J_{\rm AB} \else $J_{\rm AB}$\fi}
\newcommand{\hab}{\ifmmode H_{\rm AB} \else $H_{\rm AB}$\fi}
\newcommand{\kab}{\ifmmode K_{\rm AB} \else $K_{\rm AB}$\fi}

\newcommand{\jveg}{\ifmmode J_{\rm Vega} \else $J_{\rm Vega}$\fi}
\newcommand{\hveg}{\ifmmode H_{\rm Vega} \else $H_{\rm Vega}$\fi}
\newcommand{\kveg}{\ifmmode K_{\rm Vega} \else $K_{\rm Vega}$\fi}

\newcommand{  \Chisq    }{\ifmmode \chi^{2} \else $\chi^{2}$}
\newcommand{  \nelec    }{\ifmmode n_{e} \else $n_{e}$\fi}     
\newcommand{  \nh       }{\ifmmode n_{\rm H} \else $n_{\rm H}$\fi}     
\newcommand{  \Ncol     }{\ifmmode N_{\rm col} \else $N_{\rm col}$\fi} 
\newcommand{  \NH       }{\ifmmode N_{\rm H} \else $N_{\rm H}$\fi}     

\def\sun{\hbox{$\odot$}}
\def\ion#1#2{#1$\;${\small\rm\@Roman{#2}}\relax}

\newcommand{\kbol}{\ifmmode \kappa_{\rm bol} \else $\kappa_{\rm bol}$\fi}

\newcommand{\NeV}{\ifmmode \left[{\rm Ne}\,\textsc{v}\right]\,\lambda3427 \else [Ne\,\textsc{v}]\,$\lambda3427$\fi}
\newcommand{\NeIII}{\ifmmode \left[{\rm Ne}\,\textsc{iii}\right]\,\lambda3870 \else [Ne\,\textsc{iii}]\,$\lambda3870$\fi}
\renewcommand{\neiii}{\ifmmode \left[{\rm Ne}\,\textsc{iii}\right] \else [Ne\,\textsc{iii}]\fi}

\newcommand{\Lbat  }{\ifmmode L_{\rm BAT} \else $L_{\rm BAT}$\fi}
\newcommand{\lamEdd}{\ifmmode \lambda_{\rm Edd} \else $\lambda_{\rm Edd}$\fi}

\newcommand{\lognh}{\ifmmode \log(\NH/\cmii) \else $\log(\NH/\cmii)$\fi}

\newcommand{\fnev }{\ifmmode F_{\nev} \else $F_{\nev}$\fi}
\newcommand{\Lnev }{\ifmmode L_{\nev} \else $L_{\nev}$\fi}
\newcommand{\ewnev}{\ifmmode {\rm EW}_{\nev} \else EW$_{\nev}$\fi}

\newcommand{\fneiii }{\ifmmode F_{\neiii} \else $F_{\neiii}$\fi}
\newcommand{\foiii}{\ifmmode F_{\oiii} \else $F_{\oiii}$\fi}
\newcommand{\Loiii}{\ifmmode L_{\oiii} \else $L_{\oiii}$\fi}
\newcommand{\fhb}{\ifmmode F_{\hb} \else $F_{\hb}$\fi}
\newcommand{\ohb}{\ifmmode \oiii/\hb \else $\oiii/\hb$\fi}

\newcommand{\fdet}{\ifmmode f_{\rm det} \else $f_{\rm det}$\fi}

\newcommand{\ffx}{\ifmmode \fnev/\Fuhard \else $\fnev/\Fuhard$\fi}

\newcommand{\Nobj}{two}

\newcommand{\objectone}{NG-GS-10013609}
\newcommand{\objone}{GS-10013609}
\newcommand{\objecttwo}{GN~42437}
\newcommand{\objtwo}{GN~42437}

\newcommand{\mbhmin}{\ifmmode M_{\rm BH,min} \else $M_{\rm BH,min}$\fi}
\newcommand{\mbhmax}{\ifmmode M_{\rm BH,max} \else $M_{\rm BH,max}$\fi}

\received{2025 July 14}
\revised{2025 September 7, 26}
\accepted{2025 September 28}
\submitjournal{ApJL}

\shorttitle{$z>5$ AGN with \NeV\ emission}
\shortauthors{Trakhtenbrot et al.}
\graphicspath{{./}{figures/}}

\begin{document}

\title{Insights for Early Massive Black Hole Growth from JWST Detection of the \NeV\ Emission Line}

\correspondingauthor{Benny Trakhtenbrot}
\email{bennyt@tauex.tau.ac.il}

\author[0000-0002-3683-7297]{Benny Trakhtenbrot}
\affiliation{School of Physics and Astronomy, Tel Aviv University, Tel Aviv 69978, Israel}
\affiliation{Max-Planck-Institut f{\"u}r extraterrestrische Physik, Gie\ss{}enbachstra\ss{}e 1, 85748 Garching, Germany}
\affiliation{Excellence Cluster ORIGINS, Boltzmannsstra\ss{}e 2, 85748, Garching, Germany}

\author[0000-0001-5231-2645]{Claudio Ricci}
\affiliation{Department of Astronomy, University of Geneva, 1290 Versoix, Switzerland}
\affiliation{Instituto de Estudios Astrof\'isicos, Facultad de Ingenier\'ia y Ciencias, Universidad Diego Portales, Av. Ej\'ercito Libertador 441, Santiago, Chile} 

\author[0000-0001-7568-6412]{Ezequiel Treister}
\affiliation{Instituto de Alta Investigaci{\'o}n, Universidad de Tarapac{\'a}, Casilla 7D, Arica, Chile}

\author[0000-0002-7998-9581]{Michael J. Koss}
\affiliation{Eureka Scientific, 2452 Delmer Street, Suite 100, Oakland, CA 94602-3017, USA} 
\affiliation{Space Science Institute, 4750 Walnut Street, Suite 205, Boulder, CO 80301, USA}

\author[0000-0002-7962-5446]{Richard Mushotzky}
\affiliation{Department of Astronomy, University of Maryland, College Park, MD 20742, USA}

\author[0000-0002-5037-951X]{Kyuseok Oh}
\affiliation{Korea Astronomy \& Space Science Institute, 776, Daedeokdae-ro, Yuseong-gu, Daejeon 34055, Republic of Korea}

\author[0000-0003-2196-3298]{Alessandro Peca}
\affiliation{Eureka Scientific, 2452 Delmer Street, Suite 100, Oakland, CA 94602-3017, USA}
\affiliation{Department of Physics, Yale University, P.O. Box 208120, New Haven, CT 06520, USA}

\author[0000-0002-8686-8737]{Franz E. Bauer}
\affiliation{Instituto de Alta Investigaci{\'o}n, Universidad de Tarapac{\'a}, Casilla 7D, Arica, Chile}

\author[0009-0007-9018-1077]{Kriti Kamal Gupta}
\affiliation{STAR Institute, Li\`ege Universit\'e, Quartier Agora - All\'ee du six Ao\^ut, 19c B-4000 Li\`ege, Belgium}
\affiliation{Sterrenkundig Observatorium, Universiteit Gent, Krijgslaan 281 S9, B-9000 Gent, Belgium}

\author{Tomer Reiss}
\affiliation{School of Physics and Astronomy, Tel Aviv University, Tel Aviv 69978, Israel}



\begin{abstract}
We use the narrow \NeV\ emission line detected in the recently published JWST spectra of \Nobj\ galaxies, at $z\simeq6.9$ and $5.6$, to study the key properties of the active galactic nuclei (AGN) and the supermassive black holes (SMBHs) in their centers. 
Using a new empirical scaling linking the \nev\ line emission with AGN accretion-driven (continuum) emission, derived from a highly complete low-redshift AGN sample, we show that the \nev\ emission in the \Nobj\ $z>5$ galaxies implies total (bolometric) AGN luminosities of order $\Lbol\approx(4-8)\times10^{45}\,\ergs$. 
Assuming that the radiation emitted from these systems is Eddington limited, the (minimal) black hole masses are of order $\mbh\gtrsim10^{7}\,\Msol$. 
Combined with the published stellar masses of the galaxies, estimated from dedicated fitting of their spectral energy distributions, the implied BH-to-stellar mass ratios are of order $\mbh/\mhost \approx 0.1-1$. This is considerably higher than what is found in the local Universe, but is consistent with the general trend seen in some other $z\gtrsim5$ AGN. 
Given the intrinsic weakness of the \nev\ line and the nature of the \nev--to--\Lbol\ scaling, any (rare) detection of the \NeV\ line at $z>5$ would translate to similarly high AGN luminosities and SMBH masses, thus providing a unique observational path for studying luminous AGN well into the epoch of reionization, including obscured sources.
\end{abstract}


\section{Introduction} 
\label{sec:intro}

Quasars detected at $z\gtrsim6$, which now total hundreds of sources, have challenged our understanding of supermassive black hole (SMBH) formation and early growth, inspiring models in which SMBHs grow from massive seeds and/or through exceptionally efficient, super-Eddington accretion \cite[e.g.,][and references therein]{Inayoshi2020,Fan2023}.
However, until recently, there were relatively few observations of lower-luminosity and/or obscured active galactic nuclei (AGN) at these early epochs, although such sources are expected to be (intrinsically) far more abundant \cite[e.g.,][]{Ni20,Gilli2022,Peca2023,Schindler2023}.

The James Webb Space Telescope (JWST) is revolutionizing our ability to study high-redshift AGN and SMBHs. Not only has it provided measurements of the host galaxies and larger-scale environments of previously known quasars \cite[e.g.,][]{Ding23,Yue24}, but crucially has detected several types of AGN that were previously unseen at $z\gtrsim5$. 
These include: 
broad-line AGN (candidates) now detected out to $z\gtrsim10$ \citep{Kokorev23} and to considerably lower luminosities and/or BH masses compared with the luminous quasars probed before \citep{Harikane23,Kocevski23_CEERS,Ubler23,Furtak24,Lin24,Taylor25,Juodzbalis25};
an abundant population of so-called ``little red dots''---many of which appear to harbor AGN based on the presence of broad emission lines \cite[albeit with properties that are still under intense discussion; e.g.,][]{Greene24,Matthee24};
and several obscured AGN identified either through multiwavelength emission \cite[e.g.,][]{Bogdan24,Lambrides24} or narrow emission lines of highly ionized species \cite[e.g.,][]{C24,S25,Mazzolari2025};
Many of these systems exhibit BH-to-host masses that are significantly higher than what is seen in the local Universe, reaching $\mm\approx0.1$ \cite[e.g.,][]{Maiolino23,Harikane23,Ubler23,Furtak24,Juodzbalis24,Juodzbalis25}.
These discoveries motivated an intensive discussion regarding the origin and (surprisingly high) abundance of such systems. 

Here, we study \Nobj\ star-forming, low-mass galaxies that were recently claimed to host AGN based on emission lines measured from JWST spectroscopy (\citealt[S25 hereafter]{S25}; \citealt[C24 hereafter]{C24}). 
We argue that these galaxies harbor intrinsically luminous AGN based on the published measurements of the \NeV\  emission line. 
This line is commonly used as a robust AGN tracer \cite[e.g.,][]{Zakamska2003,Gilli2010,Yuan2016,Vergani2018}, as it requires an ionizing source reaching $\gtrsim100$ eV---much higher than what stellar populations can typically provide (e.g., \citealt{Satyapal2021,Cleri2023_diag}, and references therein, for detailed discussion).
We describe the data available for the \Nobj\ galaxies in Section~\ref{sec:obs_data}. 
In Section~\ref{sec:agn_lum} we derive estimates of the intrinsic, accretion-powered luminosities of the AGN that drive the high ionization line emission in these galaxies. 
We discuss the implications for early SMBHs and their host galaxies in Section~\ref{sec:mm}, before concluding in Section~\ref{sec:summary}.
All the luminosities mentioned in this Letter are (re)calculated assuming a flat $\Lambda$CDM cosmology, with $H_0 = 70\,\Hubble$, $\Omega_\Lambda = 0.7$ and $\Omega_{\rm M}=0.3$.

\defcitealias{R25}{R25}
\defcitealias{S25}{S25}
\defcitealias{C24}{C24}

\newcommand{\papnev}{\citetalias{R25}}
\newcommand{\papone}{\citetalias{S25}}
\newcommand{\paptwo}{\citetalias{C24}}


\section{Two AGN-hosting galaxies at $z\protect\sim5.5-7$}
\label{sec:obs_data}

\begin{deluxetable*}{lccc}
\label{tab:params}
\tablecaption{Properties of the two AGN}
\tablewidth{0.475\textwidth}
\tablehead{
\colhead{Property} & \colhead{\objone} & \colhead{\objtwo} & \colhead{Reference\tablenotemark{$a$}}}
\startdata
Key spec. reference             &	\citetalias{S25}      &	\citetalias{C24}  &		\\
$\alpha$(J2000.0) [d]           &	53.117303             &	189.17219         &	1	\\
$\delta$(J2000.0) [d]           &	$-$27.764082          &	$+$62.30564       &	1	\\
$z$	                            &	6.931                 &	5.58724           &	1	\\
$D_{\rm L}$ [Gpc]               &	68.236                &	53.128            &	2	\\
$M_{\rm UV}$ [mag]              &	$-$18.7               &	$-$19.1           &	1	\\
$\log\mstar$ [\Msol]            &	$7.7^{+0.06}_{-0.07}$ &	$7.9\pm0.2$	      &	1	\\
\hline							
\fnev\   [$10^{-19}\,\ergcms$]	&	$ 3.37\pm0.75$        &	$  2.35\pm0.35$	  &	1	\\
\fneiii\ [$10^{-19}\,\ergcms$]  &   $ 3.46\pm0.36$        & $  9.0\pm0.7$     & 1   \\
\fhb\    [$10^{-19}\,\ergcms$]	&	$ 5.83\pm0.27$	      &	$ 20.2\pm0.7$     & 1 \tablenotemark{$b$}	\\
\foiii\  [$10^{-19}\,\ergcms$]	&	$33.04\pm0.33$	      &	$115.3\pm3.1$     &	1	\\
\hline							
$\log\Lnev$  [\ergs]            &	41.27	              &	40.90	          &	2	\\
$\log\Loiii$ [\ergs]            &   42.27                 &	42.59             &	2	\\ 
\hline							
$\log\Lbol(\Lnev)$	[\ergs]     &	45.9	&	45.6	&	Eq.~\ref{eq:LNeV_Lbol_scale}, \papnev\,	\\
$\log\Lbol(\Loiii)$	[\ergs]     &	45.8	&	46.1	&	Eq.~\ref{eq:Loiii_Lbol_H04}, \cite{Heckman04}	\\
\hline							
$\log\mbhmin$	[\Msol] 	    &	7.75	&	7.38	&	Eq.~\ref{eq:LNeV_Lbol_scale} \& $\Lbol\leq\Ledd$\\
$\mbhmin/\mstar$	            &	$\approx1.1$	&	$\approx0.32$	&	1, 2\\
\enddata
\tablenotetext{$a$}{References for tabulated properties: (1) - the relevant original papers: \papone\ and/or \cite{DEugenio25_JADES_DR3} for \objone, and \paptwo\ for \objtwo; (2) - this work.}
\tablenotetext{$b$}{For \objtwo, \hb\ was not measured directly, so instead we list the \fhb\ derived by combining the \foiii\ and the \ohb\ values tabulated in \paptwo.} 
\end{deluxetable*}


This work focuses on \Nobj\ galaxies with narrow \NeV\ emission lines identified in recently published JWST/NIRSpec-based studies.
Table~\ref{tab:params} lists the key properties of the two galaxies -- including those reported in the papers presenting their \nev\ measurements (\citetalias{C24},\citetalias{S25}), and those derived throughout the present study.
We stress that the \Nobj\ studies that presented the galaxies have already analyzed their spectra and provided significant evidence for their AGN nature, based on their luminous \nev\ line emission. Below we revisit some of these arguments and further strengthen this basic conclusion.

These \Nobj\ galaxies are the only $z>5$ cases we are aware of where the robust detection and measurement of the \nev\ line were made public. Given that these are drawn from large samples of hundreds of high-redshift galaxies with JWST spectroscopy, they are likely {\it not} representative of the overall galaxy population.
Recently, two additional galaxies were claimed to present some \nev\ emission, as reported by \cite[][GS-z9-0 at $z=9.43$]{Curti25} and by \cite[][GS-81034 at $z = 5.39$]{Tang25}.
In both these cases, the \nev\ line is only marginally detected, with  a signal-to-noise ratio of $S/N\simeq3$ (varying across reduction methods).
We therefore prefer not to include these galaxies in our study (but see possible implications mentioned in Section~\ref{sec:summary}).


\subsection{\objectone}

\objectone\ (\objone\ hereafter), at $z=6.931$, is part of a sample of 42 galaxies with AGN emission line signatures presented by \papone, based on JWST/NIRSpec observations carried out as part of the JWST Advanced Deep Extragalactic Survey \citep{Eisenstein2023_JADES_overview}. 
The narrow ($\fwhm\simeq215\,\kms$) \NeV\ line is robustly detected, with a luminosity corresponding to  $\Lnev=1.9\times10^{41}\,\ergs$.
Other lines of interest that are robustly detected  include \CIII, \NeIII, \hbeta, and \OIII\ \cite[see Table~\ref{tab:params};][]{DEugenio25_JADES_DR3}.
The basic galaxy properties were derived by fitting the NIRSpec/PRISM spectra that are part of JADES, using the \texttt{BEAGLE} code \citep{Chevallard16}. This analysis yielded a stellar mass of $\log(\mstar/\Msol)=7.7^{+0.06}_{-0.07}$ and a star formation rate of ${\rm SFR}\approx 4\,\mpyr$.
\cite{Maiolino25_Xray} derived a $3\sigma$ upper limit on the (rest-frame) 2--10\,\kev\ luminosity of $L(2-10\,\kev)< 5.3\times10^{42}\,\ergs$.

As for the AGN nature of \objone, we first note that \papone\ classified it as an AGN (and thus included it in their sample), based on the mere detection of \NeV.\footnote{This is the only source in the \papone\ sample where \NeV\ is robustly detected.} 
Further evidence for an AGN in this source comes from its location in the \ciii/\HeIIuv\ vs.\ \nev/\ciii\ diagnostic diagram \cite[Fig.~5 in \papone; following][]{Feltre16}, as well as in the \oiii/\hb\ vs.\ \nev/\neiii\ diagnostic diagram put forward by \cite{Cleri2023_diag}. 
The latter is shown in Figure~\ref{fig:line_ratios}, illustrating how most low redshift AGN from the BASS/DR2 catalog \citep{Koss2022_DR2_overview,Oh2022_DR2_NLR} fall in the ``AGN'' region. 
Finally, the high \Lnev\ itself provides further support for an AGN, as we discuss below.
%

\subsection{\objecttwo}

\objtwo\ is a $z=5.58724$ galaxy for which deep JWST/NIRSpec spectroscopy was obtained as part of a study of several epoch-of-reionization galaxies in the GOODS-North (GN) field (JWST Project ID: 1871, PI: Chisholm). 
The exquisite NIRSpec spectrum of \objtwo\ revealed many emission lines, including \nev, \neiii, \oiii, and \HeIIop. 
Specifically, the narrow ($\fwhm\simeq174\,\kms$) \NeV\ line has a luminosity of $\Lnev=7.9\times10^{40}\,\ergs$.\footnote{We rely on the observed line fluxes from \paptwo, uncorrected for attenuation, so to be consistent with what is available for \objone.}
The deep JWST/NIRCam imaging available from the First Reionization Epoch Spectroscopically Complete Observations project \cite[FRESCO;][]{Oesch23_fresco}, combined with past HST imaging and using the \texttt{BAGPIPES} spectral fitting code \citep{Carnall18_bagpipes}, yielded $\log(\mstar/\Msol)=7.9\pm0.2$ and ${\rm SFR}\simeq 10\,\mpyr$.
\paptwo\ further reported an upper limit on the X-ray emission of $L(2-10\,\kev)< 4.5\times10^{43}\,\ergs$.

\begin{figure}
    \centering
    \includegraphics[width=0.475\textwidth]{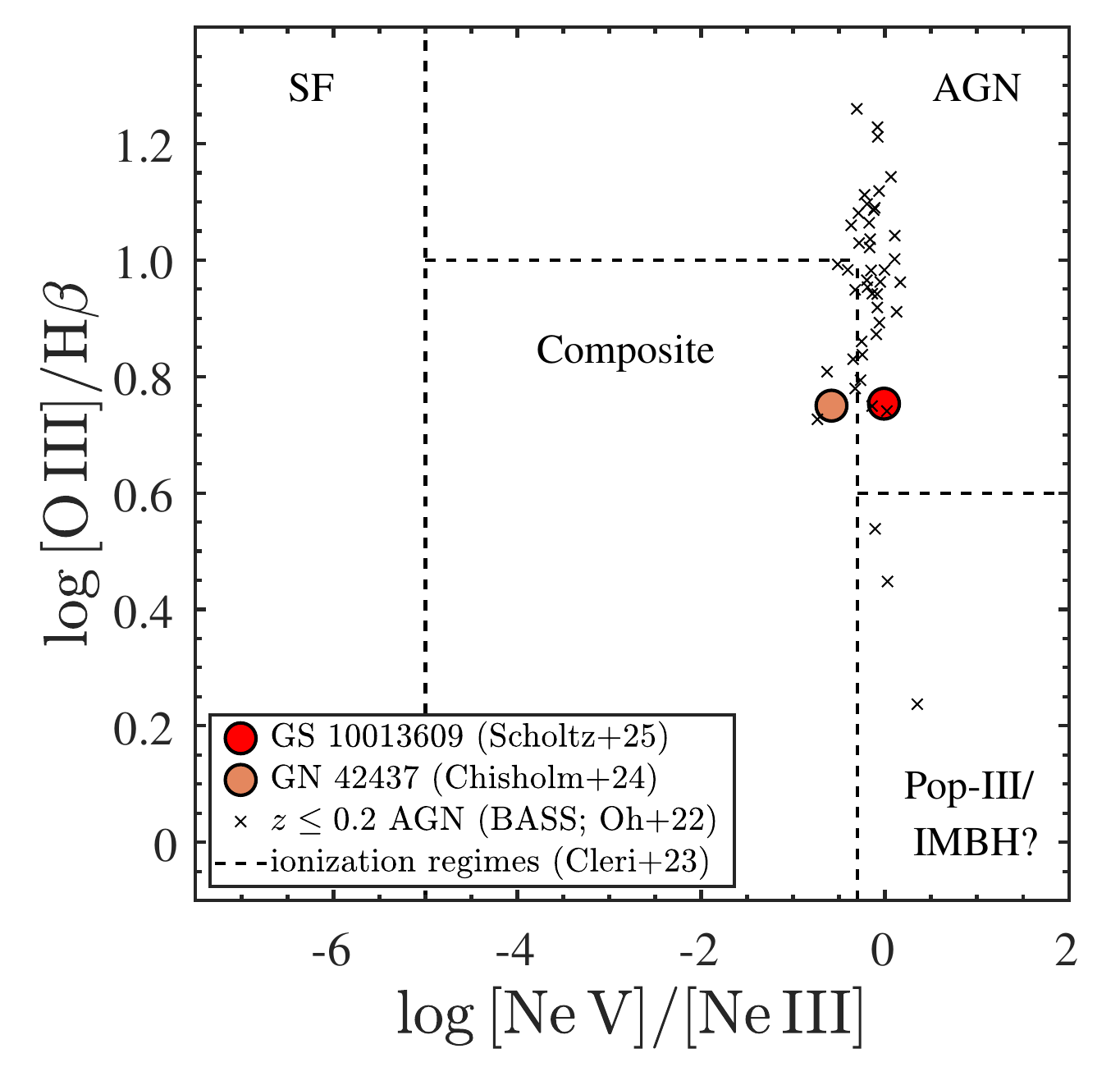}
    \caption{The \NeV/\NeIII\ vs. \OIII/\hb\ line ratio diagnostic diagram, with regions dominated by various ionization mechanisms marked following \cite{Cleri2023_diag}. 
    The \Nobj\ galaxies under study (large colored symbols) are consistent with AGN-driven photoionization, albeit with some contribution from other sources for \objtwo. 
    Low redshift, bona fide narrow-line AGN (black crosses; drawn from BASS/DR2, \citealt{Oh2022_DR2_NLR}) have similar line ratios to those of the \Nobj\ high-redshift galaxies we study here. 
    We argue that the high \nev\ line luminosity, that is \Lnev\ itself, should also be considered when interpreting the ionization mechanism.}
    \label{fig:line_ratios}
\end{figure}

\paptwo\ employed photoionization modeling and line ratio diagnostic diagrams, concluding that the emission lines in \objtwo\ can only be explained with a dominant BH-accretion-based ionizing source. 
However, a significant contribution from stars, perhaps as much as $\approx2/3$ of the {\it hydrogen} ionizing radiation, cannot be ruled out. The potentially ``composite'' nature of \objtwo\ can be demonstrated by the \nev/\neiii\ vs.\ \oiii/\hb\ diagnostic diagram in Fig.~\ref{fig:line_ratios}. We note that \objtwo\ is located very close to the pure AGN region in that diagram, where we also find a nonnegligible fraction ($\sim$16\%) of the low-redshift, bona fide BASS AGN.

Importantly, we argue that the \nev\ line \emph{luminosity} (i.e., \Lnev), and not just line \emph{ratios}, provides further support for an AGN in this system.
Specifically, if most of the high \nev\ luminosity, corresponding to roughly $\approx5\times10^7\,L_\odot$, is not due to an AGN, it implies an exceptionally large (and/or massive) population of Pop-III stars and/or intermediate mass black holes (IMBHs), the existence of which cannot be further supported by the data in hand.
As a reference for the high \Lnev\ of \objtwo\ we mention two recent studies of this line in low-redshift galaxies.
First, over 80\% of the $\sim$150 \nev-detected sources in the recently published study of \NeV\ emission in BASS AGN \citep[][R25 hereafter]{R25} have \Lnev\ that is lower than that of \objtwo.
Second, among the 60 \nev-emitting galaxies in the Coronal Line Activity Spectroscopic Survey \cite[CLASS;][]{Reefe23_CLASS_cat}---all of which can be independently classified as AGN---about 13\% have \Lnev\ that is lower than that of \objtwo. Moreover, all the \nev-emitting CLASS AGN are found in hosts with $\mstar\gtrsim5\times10^{9}\,\Msol$, that is $\gtrsim2$\,dex higher than what is estimated for \objtwo.
Thus, even for those systems in the local Universe where \nev\ is as luminous as, or lower than, what is seen in \objtwo, if we were to suspect a significant non-AGN contribution to \Lnev, this would correspond to galaxies with significantly larger populations of (young) stars than the data for \objtwo\ allow.\footnote{We note that the \Lnev\ of \objone\ is similar to the median value among the CLASS AGN.}


\section{AGN energetics}
\label{sec:agn_lum}

%
We next estimate the bolometric AGN luminosities, \Lbol, of the \Nobj\ galaxies under study.
We focus on the recently published scaling between \Lnev\ and the (intrinsic), accretion-driven, AGN radiation, as traced by the ultrahard X-ray luminosity in the $14-195\,\kev$ band, \Luhard, which was described in \papnev.
The \papnev\ study is based on a large sample of low-redshift, ultrahard X-ray selected narrow-line AGN from BASS/DR2 \citep{Koss2022_DR2_overview}, which have high-quality optical spectroscopy. The ultrahard X-ray band used to define the sample and to link \nev\ line emission with AGN continuum radiation is essentially unaffected by obscuration.
The specific scaling suggested in \papnev, $\log\Luhard=\log\Lnev + 3.75$, represents the median value found among the \nev-detected sources in the \papnev\ sample, with no correction of the \nev\ emission for source dust attenuation, and has a scatter of 0.45\,dex. 
Combining this scaling with a simple, universal bolometric correction of $\Lbol=8\times\Luhard$ (see also \citealt{Koss2022_DR2_overview,Gupta24}), \papnev\ conclude that \Lbol\ can be estimated directly from \Lnev\ following:
\begin{equation}
    \log\Lbol=\log\Lnev + 4.65 \, .
    \label{eq:LNeV_Lbol_scale}
\end{equation} 
All of the \nev-detected narrow-line AGN in \papnev\ have $\log(\Lbol/\Lnev) > 3.3$ and only $<5\%$ ($<18\%$) have $\log(\Lbol/\Lnev) < 3.65$ ($4.15$, respectively). 
Also, the \nev\ line luminosities and widths of the \Nobj\ galaxies under study are well within the range covered by the \papnev\ sample.
Therefore, the scaling we adopt here should not be considered biased (high).
We verified that using the more elaborate, nearly linear relation linking \Lnev\ and continuum AGN emission presented in \papnev\ (their Eq.~1), or their $\nev-\Lhard$ scaling, would not affect our results. 
Given the large scatter in all these scaling relations, we prefer to use the former, simpler Eq.~\ref{eq:LNeV_Lbol_scale}.
With this scaling, we obtain $\Lbol \simeq 8.5\times10^{45}$ and $3.6\times10^{45}\,\ergs$, for \objone\ and \objtwo\ (respectively).
AGN bolometric corrections have a significant scatter, and may also vary with certain AGN properties \cite[e.g.,][and references therein]{Vasudevan07,Duras20,Gupta24}. 
However, using more elaborate bolometric corrections is not warranted here, given the kind of approximate argumentation we pursue. 
Combining the 0.45\,dex scatter on the scaling between \nev\ and AGN continuum emission (\papnev) and the scatter on bolometric corrections deduced from the BASS sample, of 0.4\,dex \citep{Gupta24}, the uncertainty on our \nev-based estimates of \Lbol\ is $\sim$0.6\,dex.

To illustrate the effect of relying on other narrow emission lines from high ionization species on our assessment of AGN energetics in these \Nobj\ galaxies, we use scaling relations suggested for the \OIII\ line. 
The large SDSS study of \cite{Heckman04} suggested 
\begin{equation}
    \log\Lbol = \log\Loiii + 3.54 \,.
    \label{eq:Loiii_Lbol_H04}
\end{equation} 
The X-ray informed study of \cite{Heckman05} reported a scatter of $\approx$0.5 dex on the scaling between \Loiii\ and AGN X-ray emission, however more recent studies found that this scatter could be larger, exceeding 0.6 dex \cite[e.g.,][]{Berney2015,Ueda2015_OIII,Oh2022_DR2_NLR}. This is expected, as \oiii\ may be contaminated by galaxy-scale star forming regions and/or reflect past episodes of SMBH accretion. 
Notwithstanding these limitations,  Eq.~\ref{eq:Loiii_Lbol_H04} yields $\Lbol(\oiii)\approx6.3\times10^{45}$ and $1.4\times10^{46}\,\ergs$, for \objone\ and \objtwo\ (respectively). 
These differ from the \nev-based estimates of \Lbol\ by $-0.13$ and $+0.55$\,dex (respectively). 
If we instead use the $\oiii+\hb$ prescription of \citet[Eq.~1 there]{Netzer2009}, we obtain \Lbol\ estimates that differ from our \nev-based estimates by either $-0.6$\,dex (for \objone), or by $+0.1$\,dex (for \objtwo).
For reference, we note that the studies presenting the \Nobj\ sources have also provided rough estimates of \Lbol, based on the \oiii\ and \hb\ lines.
\papone\ estimated $\Lbol\simeq10^{44}\,\ergs$ for \objone, which is significantly lower than our \nev-based estimate (by 2\,dex; citing Hirschmann et al.\ in prep.). 
For \objtwo, \paptwo\ estimated \Lbol\ to be as high as $\Lbol\simeq10^{45}\,\ergs$ based on the \cite{Netzer2009} prescription(s), but assuming only $1/3$ of \Loiii\ is due to an AGN. 
This is in excellent agreement with our \nev-based \Lbol\ estimate, if we instead assume that 100\% of \Loiii\ is due to an AGN.

In what follows, we focus on \Lnev\ and Eq.~\ref{eq:LNeV_Lbol_scale} as the primary method to estimate the bolometric AGN luminosities for the \Nobj\ galaxies. This is motivated by the higher ionization potential needed for \nev\ (compared with \oiii), making it a high-purity AGN tracer, as well as the possibly closer distance of the \nev-emitting gas to the ionizing AGN central engine. 
We note that the metallicity in the line-emitting regions of the $z>5$ galaxies under study may be lower than in the low-redshift samples used to calibrate the aforementioned scaling relations, as suggested by various analyses of $z>5$ galaxies with JWST spectroscopy \cite[see, e.g.,][and references therein]{Stark2025_rev}. 
However, a lower metallicity would result in a {\it higher} $\Lbol/L_{\rm line}$, and so would imply a yet higher \Lbol\ given the measured \Lnev\ (or \Loiii) we use. 
This is exemplified in the works by \cite{Cleri2023_diag} and \cite{McKaig2024}, which looked specifically into how \NeV\ line emission may be affected by various properties of the ionized gas and incident radiation, including low metallicity.

Our main conclusion therefore is that the luminous narrow \nev\ line emission observed in \objone\ and \objtwo\ implies that they highly likely harbor obscured (i.e., ``Type-II'') AGN with bolometric luminosities of order $\Lbol\gtrsim3\times10^{45}\,\ergs$. This is higher than what was deduced by the previous studies of these galaxies (\papone\ and \paptwo).

Given the high \Lbol\ we infer from \nev, the upper limits on X-ray emission from the \Nobj\ galaxies imply extremely high bolometric--to--X-ray radiation ratios, $\Lbol/\Lhard\gtrsim80$ for \objtwo\ and $\gtrsim1600$ for the higher-\Lnev\, \objone.
These are (far) higher than what is seen in large samples of lower redshift (X-ray selected) AGN, where $\Lbol/\Lhard\lesssim50$ \cite[e.g.,][]{Duras20,Gupta24}. 
At face value, the implied X-ray weakness of the \Nobj\ AGN we study is in line with what is seen in other samples of JWST-detected $z>5$ AGN \cite[][]{Ananna24,Yue24_LRD_Xrays,Maiolino25_Xray}, which has been interpreted to be caused either by significant obscuration by dense circumnuclear gas \cite[e.g.,][]{Maiolino25_Xray} or by X-ray weak SEDs originating from super-Eddington accretion flows \cite[e.g.,][]{Lambrides2024_superEdd}.
In this context, our \Nobj\ AGN are clearly capable of producing copious amounts of $>$0.1 keV radiation, given their intense \nev\ emission, which could in principle be interpreted as further support of high line-of-sight obscuration toward the central X-ray source.
We caution, however, that the X-ray weakness of our \Nobj\ AGN, and other JWST-detected sources targeted by Chandra, concerns the rest-frame ultrahard X-ray radiation (rest-frame $>$10\,\kev), while the spectral regime probed (indirectly) by the \nev\ emission is much softer. Interpreting the strong line emission and weak X-ray continuum emission of our AGN thus requires a more detailed spectral analysis, which is beyond the scope of the present {\it Letter} \cite[see, e.g.,][for a specific example spectral model for \nev-weak AGN]{Lambrides2024_superEdd}.

\section{Implications for early SMBH growth}
\label{sec:mm}

What sort of SMBHs could be driving the luminous AGN in the \Nobj\ galaxies we study, which seem to have $\Lbol\gtrsim3\times10^{45}\,\ergs$, and what might this mean for early SMBH growth and galaxy evolution?

First, assuming that the \Nobj\ sources under study radiate through Eddington-limited accretion, one can deduce a \emph{lower limit} on \mbh\ by harnessing the Eddington limit itself, namely (for solar metallicity gas): 
\begin{equation}
\mbh \gtrsim \mbhmin \equiv (\Lbol/1.5\times10^{38}\,\ergs)\,\Msun\, .
\label{eq:mbh_min}
\end{equation}
Since the gas metallicity in the \Nobj\ AGN could be lower-than-solar, we note that for pure (ionized) hydrogen gas the Eddington luminosity would be slightly lower (by $\approx$0.08 dex), and the resulting \mbhmin\ would be slightly higher (by the same amount).
For the two galaxies discussed here, Eq.~\ref{eq:mbh_min} yields $\mbh\gtrsim5.6\times10^7$ and $2.4\times10^7\,\Msun$, for \objone\ and \objtwo\ (respectively).

However, there are reasons to suspect that AGN accretion flows may sustain much higher accretion rates, above the classical Eddington limit. This is particularly the case for $z \gtrsim 5$ SMBHs, which had to grow to their high masses within a limited period \cite[see, e.g., review by][]{Inayoshi2020}. Importantly, several studies have shown that even in (extremely) super-Eddington accretion flows, i.e., when $\dot{M} >> \dot{M}_{\rm Edd}$, the emerging radiation is expected to ``saturate'' at $\Lbol\approx{\rm few}\times L_{\rm Edd}$ \cite[e.g.,][]{Abramowicz88,McKinney14,KubotaDone2019}.
If we adopt the commonly-used scaling of $\Lbol\simeq \Ledd \times (1+\ln[\dot{M}/\dot{M}_{\rm Edd}])$ \citep{Abramowicz88}, this would imply \mbhmin\ values that are \emph{lower} than those mentioned above by a factor of $\approx5.6\times$ (i.e., 0.75 dex), for accretion rates with $\dot{M}=100\times\dot{M}_{\rm Edd}$. Even in such an extreme case, the implied lower limits on \mbh\ for the two galaxies are of $\gtrsim4\times10^6\,\Msun$.

To get a sense of the \emph{maximal} BH mass, \mbhmax, we may assume that the extreme-UV emitting accretion flow that is required for producing \nev\ lines can only exist if $\lledd\gtrsim0.01$. This is indeed the range seen in many samples of persistent, radiatively efficient AGN \cite[e.g.,][and references therein]{Schulze2015,Ananna2022_XLF}.
Using this lower limit on \lledd\ would naturally yield $\mbhmax=100\times\mbhmin$, that is $\approx5.6\times10^9$ and $2.4\times10^9\,\Msun$, for \objone\ and \objtwo\ (respectively).  
The \Nobj\ \nev-detected galaxies should therefore have SMBH masses in the range of roughly $\mbh\sim(5-600)\times 10^7$ and $(2-300)\times10^7\,\Msun$, for \objone\ and \objtwo\ (respectively). 
We emphasize that these are rough estimates, which are susceptible to significant (systematic) uncertainties on the assumptions taken to derive them. 
Any variation or uncertainty on the \Lbol\ estimates 
would naturally propagate, linearly, to \mbhmin\ and \mbhmax.

The \paptwo\ study of \objtwo\ also considered several estimates of \mbh, based either on the width of (narrow) emission lines and assuming the local $\mbh-\sigs$ relation, or on their estimates of \Lbol\ and requiring $\Lbol=\Ledd$. The former approach assumes that the $\mbh-\sigs$ relation does not evolve with redshift. The latter approach is, in principle, similar to our argumentation, however we note that the \paptwo\ estimates of \Lbol\ had to account for the potential contribution of young stars to the lower-ionization Balmer and \oiii\ lines, while our analysis relies solely on the exceptionally luminous \nev\ line, which has minimal contribution from young stars (if at all; Fig.~\ref{fig:line_ratios}).
In any case, the \paptwo\ \mbh\ estimates range between $\mbh({\rm C24})\approx10^6 -10^7\,\Msol$, which is lower than our estimates by ${\approx}0.4-1.5$\,dex.

The BH masses we derived above for the \Nobj\ galaxies under study are, by themselves, not exceptional. Indeed, many known $z>5$ quasars have masses in the range $\mbh\sim10^{8-9}\,\Msun$ \cite[][and references therein]{Fan2023}. Recent JWST discoveries pushed the lower end of the SMBH mass range among fainter broad-line AGN even further \cite[e.g.,][]{Harikane23,Maiolino23,Greene24,Taylor25}.
The \Nobj\ galaxies discussed here are, however, narrow-lined, meaning that their (rest-frame) optical SEDs are dominated by stars, and not by the AGN, providing reasonably robust estimates of stellar masses (see Table~\ref{tab:params} and the original papers \papone and \paptwo). Given the high AGN luminosities implied by the high ionization line emission, one may actually suspect these stellar masses are \emph{over}-estimated, if the (rest-frame) optical SED is indeed contaminated by some continuum AGN emission. 

\begin{figure}
    \centering
    \includegraphics[width=0.475\textwidth]{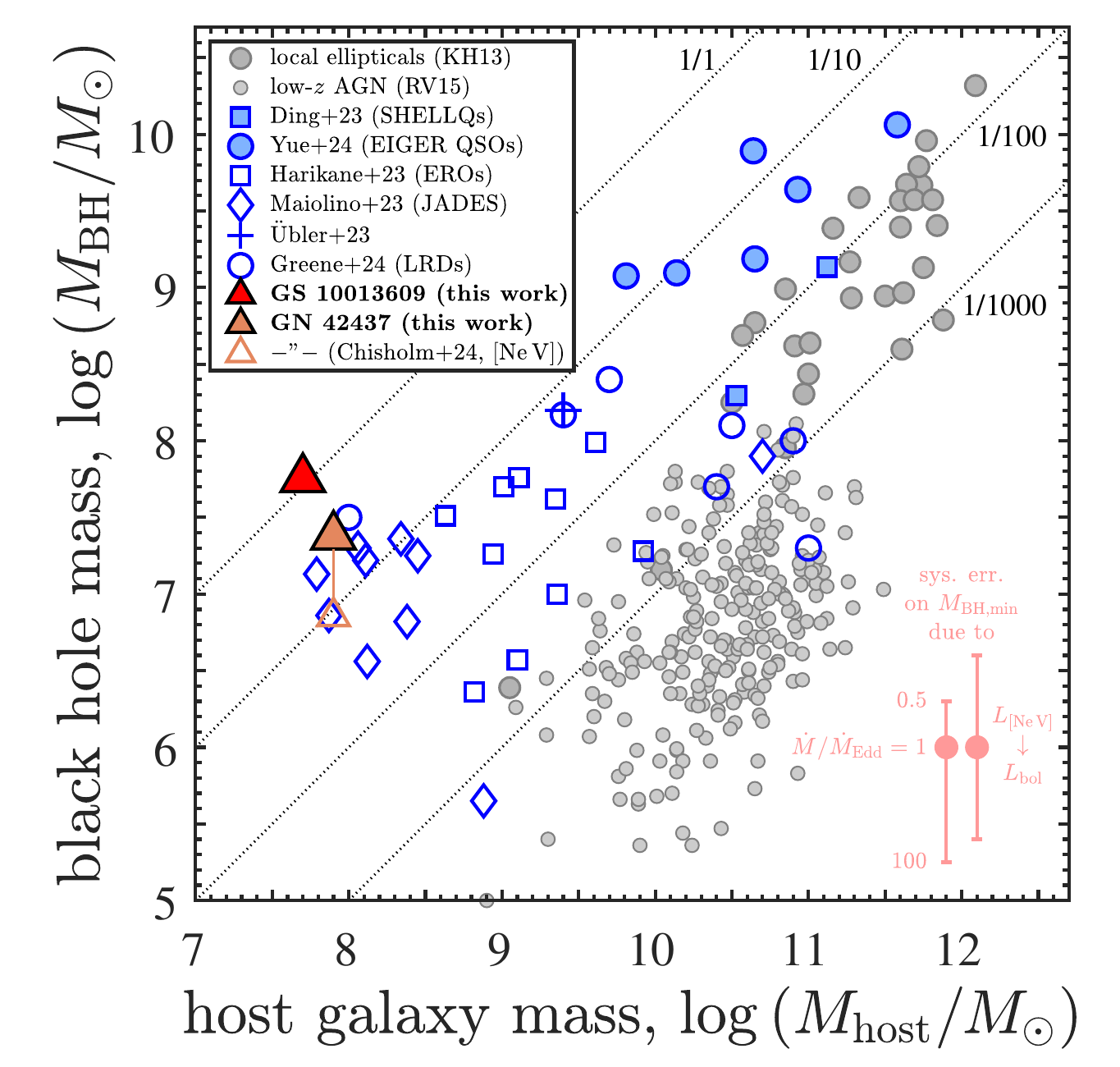}
    \caption{The SMBH--host mass plane with the \nev-based constraints for the \Nobj\ $z>5$ AGN studied here.
    We plot the two lower limits on BH masses, \mbhmin\ (colored triangles), derived by scaling the \nev\ line emission to \Lbol\ (via Eq.~\ref{eq:LNeV_Lbol_scale}) and assuming $\Lbol\leq\Ledd$.
    The systematic uncertainties associated with this scaling (0.6\,dex), and with the assumption of $\Lbol=L_{\rm Edd}$ (see text), are illustrated by the large pink marker (bottom-right corner).
    For \objtwo, we also plot an alternative estimate (lower open triangle), assuming that only $1/3$ of the ionizing radiation is due to an AGN, as suggested by \paptwo\ (a highly conservative assumption for \nev; see text).
    We include for comparison several other samples from the local and distant ($z>5$) Universe (see legend): local ellipticals (from \citealt{KH13}; large gray circles), low-$z$ broad-line AGN (from \citealt{RV15}; small gray circles); luminous $z\gtrsim6$ quasars selected from ground-based, wide-field optical-NIR surveys (from \citealt{Yue24} and \citealt{Ding23}; filled blue symbols); and several fainter, broad-line AGN identified in various spectroscopic JWST studies (from \citealt{Harikane23}, \citealt{Maiolino23}, \citealt{Ubler23}, and \citealt{Greene24}; empty blue symbols).
    Diagonal dotted lines denote BH-to-stellar mass ratios of $\mm=1:1000, 100, 10$, and $1$.
    The two sources studied here appear to have extremely high mass ratios, $\mm\approx0.1 - 1$.}
    \label{fig:mm}
\end{figure}

Figure~\ref{fig:mm} shows the \Nobj\ galaxies on the $\mbh-\mhost$ plane, in addition to several other (samples of) sources, including both relic and active systems in the local Universe, and $z\gtrsim5$ AGN of various luminosities and from different types of surveys (i.e., from large-area quasar searches and JWST deep fields).
Taking the lower values of \mbhmin\ derived above and the published \mstar\ estimates, we derive (lower limits) on the BH-to-host mass ratios of $\mm\gtrsim1.1$ and 0.3, for \objone\ and \objtwo\ (respectively).
Even if we assume that only $1/3$ of the ionizing radiation in \objtwo\ comes from an AGN, and scale down \Lbol\ and \mbhmin\ by the corresponding factor,\footnote{That is, scale the \nev-based \Lbol\ estimate down by $\times1/3$.} we obtain $\mm\approx0.1$ (see lower open triangle in Fig.~\ref{fig:mm}).
We stress that this significant scaling down of \Lbol\ represents an extremely conservative choice, given the challenge of non-AGN sources to affect the part of the ionizing radiation that drives \nev\ emission.\footnote{In other words, even if only $1/3$ of the {\it total} ionizing radiation comes from an AGN, still the vast majority of the radiation that drives \nev\ is expected to come from an AGN, and thus the \Lnev-based estimate of \Lbol\ should in principle be scaled down by a factor that is smaller than $\times 3$.}
In comparison, \paptwo\ concluded $\mm\simeq0.01$ for \objtwo, based on their lower \mbh\ estimates for that source (see their Fig.~10).

The mass ratios we derive for the \Nobj\ galaxies are obviously extremely high.
For comparison, the typical mass ratio among relic systems in the local Universe is $\mm\sim1/100-500$, and there is only limited evidence for the typical ratio to increase at high redshift, although there are a few exceptions \cite[e.g.,][and references therein]{Suh20,Ding20}. In the early Universe, $z\gtrsim5$, there are again conflicting claims for and against a significantly elevated \mm\ (compared with $z\simeq0$). Recent JWST data revealed increasingly smaller SMBHs, which appear to be ``over massive'' compared with their host galaxies \cite[c.f. the low-redshift \mm; e.g.,][]{Harikane23,Maiolino23,Furtak24,Juodzbalis24,Juodzbalis25}. 
There is, however, an ongoing debate regarding whether such elevated \mm\ ratios are representative of the intrinsic values for the underlying, not-yet-accessible population of $z\gtrsim5$ AGN (see, e.g., \citealt{Li2024_MM} vs. \citealt{Pacucci23_MM}).

Regardless of the complete distribution of \mm, every system with $\mm\gtrsim0.1$ raises the question of how the SMBH could have grown so efficiently (compared with the stellar component of its host). 
While there are several models that try to address this challenge \cite[see, e.g.,][and references therein]{Natarajan24_OBG_UZ1}, testing them would require larger samples and a better understanding of the space densities of $z>5$ AGN with certain \mbh\ and (enhanced) \mm\ ratios. 
Indeed, as further NIR spectroscopy of high-$z$ galaxies is pursued with JWST, it is possible that additional \nev-emitting AGN would be identified.
We stress, however, that the intrinsic weakness of the \nev\ line is such that any \nev-detected system would imply a rather luminous AGN.
Specifically, as noted in \papnev, \nev\ line detections down to the highest sensitivity expected for JADES would correspond to AGN luminosities of order $\gtrsim10^{45}\,\ergs$ (combining Fig.~9 in \citealt{Eisenstein2023_JADES_overview} and  Eq.~\ref{eq:LNeV_Lbol_scale}), therefore making such sources rare.
For example, if the recently reported detection of \nev\ in a $z=9.43$ galaxy \citep{Curti25} is indeed reliable,\footnote{Note that this marginal detection required 72 hours of on-source integration.}  then that source would yield $\Lbol\approx2.3\times10^{45}\,\ergs$, and thus $\mbhmin\approx1.5\times10^7\,\Msol$ and $\mm\gtrsim0.1$, merely 500 Myr after the Big Bang.

\section{Concluding remarks}
\label{sec:summary}

We have demonstrated how the mere detection of (narrow) \NeV\ line emission in spectroscopic JWST observations of $z\gtrsim5$ galaxies can provide not only a unique opportunity to uncover (obscured) AGN but also to reveal (over-)massive BHs, complementing the more common approach that focuses on broad-line AGN.
Specifically, we argue that two $z\gtrsim5$ galaxies with \nev\ detected in their JWST/NIRSpec spectra may be harboring AGN with luminosities of order $\Lbol>10^{45}\,\ergs$, based on a newly established scaling relation that links \Lnev\ to \Lbol, calibrated using a highly complete, low-redshift sample of narrow-line AGN.
We further argued that such AGN, selected solely on their \nev\ emission, could be powered by SMBHs with masses of at least $\mbh\gtrsim10^7\,\Msun$ if they are accreting at Eddington-limited rates, or larger if they their accretion rates are lower. Combined with the available estimates of the stellar masses of their hosts, we derived extremely high BH-to-host mass ratios of $\mbh/\mhost\approx0.1-1$.

The only way to avoid this latter conclusion is if  
(i) these AGN are accreting at rates significantly higher than the Eddington limit---which we think would be interesting given its potential significance to early SMBH growth and the rarity of evidence for such accretion;
(ii) their hosts include an exceptionally significant contribution from accreting IMBHs (again, a novelty; see \paptwo);
(iii) their host masses are exceptionally under-estimated---which we think is unlikely, but may have implications for other high-redshift galaxies for which \mstar\ is deduced in similar ways; 
and/or 
(iv) they are exceptionally luminous in the extreme-UV / soft X-ray regime, compared with other parts of their AGN-related SEDs (i.e., UV-optical and/or hard X-rays), in a way that is not observed in low-redshift AGN with similar high-ionization line emission.

Whatever may be the case for the \Nobj\ objects we studied, our analysis exemplifies how deep spectroscopy of galaxies in the early Universe, aiming for detecting certain (rare and weak) emission lines, can yield insights regarding the earliest phases of SMBH growth.

\acknowledgments

We thank the anonymous reviewer whose comments allowed us to improve various aspects of this {\it Letter}.
We thank R.\ Maiolino and J.\ Scholtz for providing useful information regarding \objectone, and for their invaluable comments on the manuscript. 
We also thank J.\ McKaig for his input on \nev\ emission modeling, as well as N.\ Cleri, U.\ Izotov, G.\ Mazzolari, M.\ Reefe, M.\ Volonteri, and H.\ \"Ubler for useful discussions.

B.T.\ and T.R.\ acknowledge support from the European Research Council (ERC) under the European Union's Horizon 2020 research and innovation program (grant agreement number 950533) and from the Israel Science Foundation (grant number 1849/19).
C.R.\ acknowledges support from the Swiss National Science Foundation Consolidator grant TMCG-2\_223191.
K.O.\ acknowledges support from the Korea Astronomy and Space Science Institute under the R\&D program (Project No. 2025-1-831-01), supervised by the Korea AeroSpace Administration, and the National Research Foundation of Korea (NRF) grant funded by the Korea government (MSIT) (RS-2025-00553982).
We acknowledge support from the ANID CATA-BASAL program FB210003 (E.T., F.E.B.), FONDECYT Regular grants 1241005 and 1250821 (E.T., F.E.B.), and Millennium Science Initiative, AIM23-0001 (F.E.B.). 
This research was supported by the Excellence Cluster ORIGINS which is funded by the Deutsche Forschungsgemeinschaft (DFG, German Research Foundation) under Germany's Excellence Strategy - EXC 2094 - 390783311.

B.T.\ acknowledges the hospitality of the Instituto de Estudios Astrof\'isicos at Universidad Diego Portales, the Instituto de Astrof\'isica at Pontificia Universidad Cat\'olica de Chile, and the Institut d'Astrophysique de Paris, where parts of this study have been carried out.

\bigskip
\facilities{JWST:NIRSpec}

\clearpage
\bibliography{hiz_NeV}{}
\bibliographystyle{aasjournal}

\end{document}
